# Generation of pure spin currents via nonadiabatic quantum pumping in an antiferromagnetic chain

**Leila Eslami [1*], Fatemeh Bourbour [2], Somaieh Ahmadi [3], and Santanu K. Maiti [4]**

[1] *Department of Physics, SR. C., Islamic Azad University, Tehran, Iran*

[2] *Department of Physics Education, Farhangian University, P.O. Box 889-14665 Tehran, Iran*

[3] *Department of physics, Faculty of science, Imam Khomeini International University, Qazvin, Iran.*

[4] *Physics and Applied Mathematics Unit, Indian Statistical Institute, 203 Barrackpore Trunk Road, Kolkata-700 108, India*

**E-mail:** l.eslami@srbiau.ac.ir

## Abstract

In this study, quantum spin pumping in an antiferromagnetic chain driven by time-dependent potential is investigated. The aim is to explore the possibility of generating and controlling spin currents in the absence of external bias and to examine the role of exchange field and periodic driving in the separation of spin-up and spin-down currents. The system is described using a tight-binding model, and spin-resolved currents are calculated employing the Keldysh non-equilibrium Green's function formalism. Two time-dependent potentials with a specific phase difference are applied to the two ends of the chain, while the chemical potentials of both electrodes are set equal. The results demonstrate that in the adiabatic regime (low frequencies), the response of the two spin channels is nearly identical. However, as the driving frequency increases and the system enters the nonadiabatic regime, absorption and emission processes of energy quanta become activated, leading to significant differences between spin-up and spin-down currents. The pumped current exhibits strong dependence on the chemical potential, allowing for the control of both magnitude and direction of the spin current through its adjustment. With increasing frequency, the spin current enhances, and parameters can be tuned such that the charge current nearly vanishes while a considerable spin current persists. This finding indicates the feasibility of achieving nearly pure spin pumping without net charge transfer in the antiferromagnetic chain.

The results provide a promising perspective for designing spin-pumping devices based on antiferromagnetic systems.

## 1. Introduction

In recent decades, the remarkable growth of nanotechnology and the continuous miniaturization of electronic devices have increasingly exposed the limitations of conventional charge-based technology. As device dimensions shrink to the nanoscale, phenomena such as increased energy dissipation, unwanted heat generation, current leakage, and switching-speed limitations have posed significant challenges for the development of next-generation electronic equipment [1]. Researchers are therefore seeking solutions that, while maintaining information-processing capability, reduce energy consumption and deliver faster, more stable performance. One of the most successful approaches proposed in this context is the exploitation of the electron's spin degree of freedom alongside its charge, which has given rise to a new branch of solid-state physics and nanotechnology known as spintronics [1]. In this technology, information is transmitted and processed not only by the electron's charge but also by its spin orientational feature that enables the design of devices with higher operating speeds, lower power consumption, greater storage density, and the ability to be integrated at nanoscale dimensions. For this reason, spintronics is now regarded as one of the most important research fields in condensed matter physics, and it has found wide-ranging applications in magnetic memories, spin transistors, spin filters, magnetic sensors, and quantum information processing systems [1].

The operation of any spintronic device relies on the generation, transport, and control of spin current. Unlike charge current, which arises solely from the movement of electrons, spin current represents the transfer of spin angular momentum and can occur either accompanied by a charge current or even in the absence of net charge transfer [2]. The generation of pure spin current is of practical importance, because in such a scenario the energy dissipation caused by the motion of electric charges is significantly reduced, and the resulting Joule heating in the circuit is also lowered. Hence, devising methods for the efficient generation and control of spin current has become one of the central research topics in spintronics. To date, various mechanisms have been proposed to achieve spin current, including spin injection from ferromagnetic electrodes, exploitation of spin–orbit interaction, application of external magnetic fields, and utilization of the chirality-induced spin selectivity effect [2-6].

Despite the success of these approaches, each suffers from certain limitations. The use of ferromagnetic electrodes is typically hampered by the conductivity mismatch problem at the interface between the ferromagnetic metal and the semiconductor, which severely reduces spin injection efficiency [3]. On the other hand, employing spin–orbit interaction requires materials containing heavy elements or engineered structures, and in many molecular and organic systems the strength of this interaction is insufficient to induce significant spin polarization [3]. Moreover, controlling external magnetic fields at the nanoscale is experimentally challenging; in addition to increasing fabrication complexity, it can generate unwanted magnetic fields in adjacent circuit elements [7]. These limitations have driven researchers in recent years to turn their attention toward alternative systems and mechanisms for generating and controlling spin current.

One of the most promising candidates in this regard is the use of antiferromagnetic materials [8]. Unlike ferromagnetic materials, which possess a nonzero net magnetization, in antiferromagnets the magnetic moments of the sublattices are aligned in opposite directions, so that at equilibrium the total magnetization of the system is zero. This feature eliminates stray magnetic fields and allows many devices to be integrated without inducing unwanted magnetic interactions [8]. Moreover, spin dynamics in antiferromagnetic materials occur in the terahertz frequency range, which is several orders of magnitude higher than in ferromagnets [9]. This property, together with high stability against external perturbations, good robustness against disturbing magnetic fields, and short response times, has established these materials as the next generation of active elements for spintronic devices [8,9]. In recent years, the study of spin transport, spin-transfer torque, and spin caloritronic phenomena in antiferromagnetic structures has grown considerably [8–10].

A promising approach in this area is the engineering of magnetic order in antiferromagnetic structures, which offers enhanced control over spin currents [3]. Experimental studies have shown that in such structures the spin coherence length is significantly increased and energy dissipation arising from spin–orbit interactions is reduced [3].

In addition to the choice of suitable material, the method of generating spin current is also of fundamental importance. *One attractive approach for generating current in mesoscopic systems is the use of quantum pumping.* In this phenomenon, instead of applying a direct potential difference between two electrodes, time-dependent periodic modulations of the system parameters are employed to transport carriers. If at least two independent parameters are varied periodically with an appropriate phase difference, a direct current can be generated even in the absence of any external bias [11]. This process, initially studied within the framework of adiabatic pumping, was later extended to non-adiabatic regimes and is now regarded as one of the key topics in quantum transport physics. Since this method does not require direct voltage, energy dissipation is reduced and current control through external parameters becomes possible with greater precision [11].

Numerous theoretical and experimental studies have been carried out on quantum charge pumping in various systems, including quantum dots, nanowires, graphene nanoribbons, mesoscopic rings, topological insulators, and molecular structures [11, 12]. Subsequently, the concept of quantum spin pumping also attracted attention; with an appropriate choice of time-dependent parameters, it became possible to generate a pure spin current without inducing a charge current [2, 13]. This feature makes spin pumping a suitable candidate for designing spin current sources with low energy dissipation.

In recent years, the use of one-dimensional magnetic structures and systems with magnetic order for controlling spin transport has attracted considerable attention. Magnetic chains provide a suitable platform for studying spin transport owing to their modelling simplicity, the controllability of interaction parameters, and the possibility of a detailed investigation of quantum effects [14]. Among these systems, antiferromagnetic chains are of particular importance because, despite possessing zero net magnetization, the exchange interaction between neighbouring spins can give rise to transport behaviours that differ from those in ferromagnetic systems [14]. However, many studies on spin transport in these structures have been limited to equilibrium conditions, the application of a direct bias, or magnetic excitations, and the investigation of quantum spin pumping in such systems remains very scarce [14, 15].

On the other hand, the development of theoretical methods based on the nonequilibrium Green's function (NEGF) formalism has enabled the detailed study of open quantum systems under time-dependent perturbations [12, 16]. In addition to considering the system's interaction with the electrodes, this formalism can incorporate into the transport calculations the effect of absorption and emission of energy quanta induced by the alternating field. Therefore, the NEGF method is regarded as one of the most precise tools available for investigating quantum pumping in mesoscopic structures and has been widely employed in recent years to study charge and spin transport in various systems [12, 16].

Despite significant progress in the field of quantum pumping and spin transport, the study of spin pumping in antiferromagnetic chains has not been studied yet. To date, most studies have focused on ferromagnetic systems or structures with spin–orbit interaction, and the role of antiferromagnetic order in the generation of spin current driven by time-dependent potentials has received less attention [13, 17, 18]. Moreover, it is still unclear how parameters such as the exchange field, the phase difference between alternating potentials, and the chemical potential can control the magnitude and direction of the pumped spin current. Addressing these questions is not only of fundamental importance but could also pave the way for designing low-power spin current sources without the need for external magnetic fields [7, 10].

In this work, quantum spin pumping in an antiferromagnetic chain coupled to two non-magnetic electrodes is studied. To this end, two time-dependent alternating potentials with a specific phase difference are applied to the two ends of the chain, and the resulting spin current is calculated in the absence of any external bias. The theoretical model of the system is formulated based on a tight-binding Hamiltonian, and the Keldysh nonequilibrium Green's function (NEGF) formalism is employed to investigate the time-dependent transport.

Previous theoretical studies have shown that in antiferromagnetic structures, strong exchange interactions can give rise to anomalous transport phenomena that are not achievable in ferromagnetic systems [10]. Studies on spin chains have demonstrated that the presence of antiferromagnetic order can induce significant spin polarization in response to external excitations [19]. This feature, together with the possibility of electrically controlling the spin current through applied potentials, offers a promising prospect for practical applications in spintronic devices.
The results of this study indicate that, by an appropriate choice of system parameters, a pure spin current with controllable direction and magnitude can be generated a capability that can be harnessed in the design of a new generation of spintronic devices based on antiferromagnetic materials [20–24]. Moreover, the findings of this work can contribute to a deeper understanding of the fundamental phenomena associated with spin transport in one-dimensional systems and pave the way for the design of more efficient quantum systems for spin-based information processing [25–28]. Finally, the development of such systems may represent an important step toward the realization of novel computational technologies with ultra-low power consumption and high operating speed [29].

## 2. Theoretical method

The system under consideration, as schematically depicted in Fig. 1, consists of an antiferromagnetic atomic chain comprising $\mathrm{N} = 20$ sites, confined by potential barriers at both ends $(\mathrm{i} = 1, 20)$ and coupled to two leads. To investigate the spin transport properties, time-dependent oscillating potentials with identical frequency and amplitude, but with a specified phase difference, are applied to the two ends of the chain. It is noteworthy that, to maintain equilibrium conditions and to ensure the absence of any current in the ground state, the static potential difference (time-independent bias) between the two ends of the chain is set to zero.

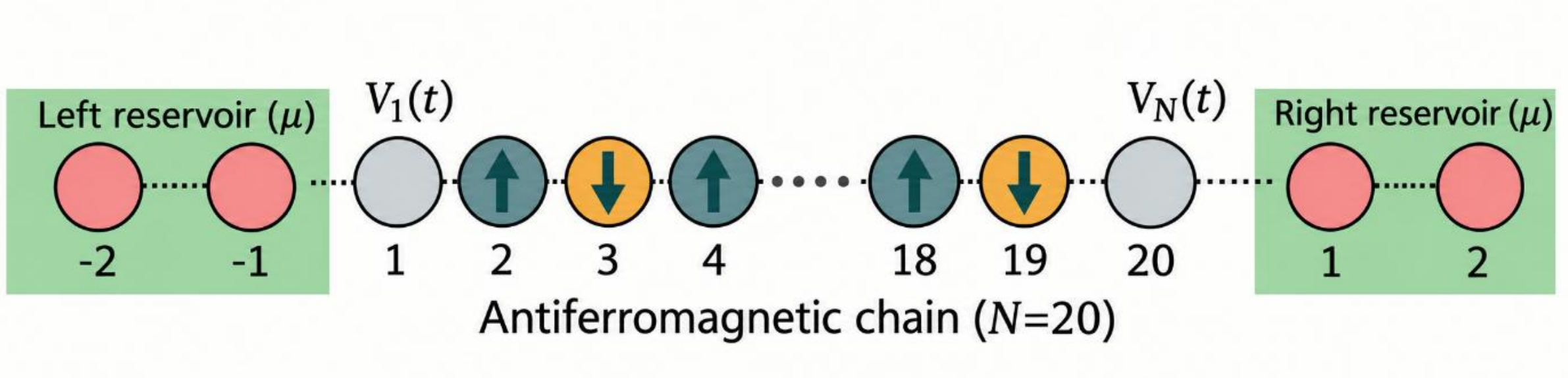

Figure 1. Schematic representation of a 20-site atomic chain with an antiferromagnetic configuration, connected to two leads with equal chemical potentials on the left and right. The atoms at the two ends of the chain are subjected to periodic potentials with a phase difference.

Under these conditions, any induced current in the chain arises only from the effects of the time-dependent potentials, and the main purpose of this study is to investigate the pumped currents in the absence of bias. The total Hamiltonian of the system is thus defined as the sum of four main terms:

$$H = H_{chain} + H_{res} + H_{cpl} + \sum_{\alpha=1}^{2} V_p \cos(\Omega_0 t + \varphi_\alpha), \quad (1)$$

where the term $H_0 = H_{chain} + H_{res} + H_{cpl}$ includes the time-independent part of the Hamiltonian. Here, $H_{chain}$ includes the antiferromagnetic exchange interactions between the localized spins in the chain, $H_{res}$ corresponds to the left and right leads, and $H_{cpl}$ describes the spin-dependent tunneling coupling between the chain and the leads. The last term represents the time-dependent oscillating potentials with identical amplitudes ($V_p$) and frequencies ($\Omega_0$), and different phases ($\varphi_1$ and $\varphi_2$), which are applied to the left and right ends of the chain ($\mathrm{i} = 1, \mathrm{N}$), respectively. To provide a more detailed description of the chain, electrodes, and the coupling between the electrodes and the chain, the corresponding Hamiltonians are defined as follows,

$$H_{chain} = \sum_{i=1,\sigma}^{N} c_{i,\sigma}^{\dagger} \varepsilon_i c_{i,\sigma} + \sum_{i=2,\sigma}^{N-1} c_{i,\sigma}^{\dagger} (h_i . \sigma) c_{i,\sigma} + \sum_{i,\sigma} t_{i,i+1} \left( c_{i,\sigma}^{\dagger} c_{i+1,\sigma} + h.c. \right) \quad (2)$$

$$H_{res} = \sum_{\alpha=L,R} \sum_{j,\sigma} \varepsilon_{j\alpha} d_{j\sigma\alpha}^{\dagger} \, d_{j\sigma\alpha} + \sum_{\alpha=L,R} \sum_{j,\sigma} t_{j,j+1\alpha} \, d_{j\sigma\alpha}^{\dagger} d_{j\alpha\sigma} \quad (3)$$

$$H_{cpl} = \sum_{\alpha=L,R} \sum_{\sigma} (w_\alpha c_{\alpha,\sigma}^{\dagger} d_{\alpha,\sigma} + H.c.). \quad (4)$$

In the above relations, the index $\sigma=\uparrow,\downarrow$ denotes the electron spin. The operators $c_{i,\sigma}^{\dagger}$ ($c_{i,\sigma}$) are the creation (annihilation) operators for an electron with spin $\sigma$ at the $i$-th site of the chain. The parameter $\varepsilon_i$ represents the onsite energy at site $i$. We take $\varepsilon_i=0$ for $\mathrm{i}=2,\ldots,\mathrm{N}-1$, and $\varepsilon_1=\varepsilon_N=E_b$ to model the confinement barriers. The term $h_i\sigma$ models the exchange interaction of the effective magnetic field with the electron spins [19], where $h_i$ is a scalar strength and $\sigma$ is the Pauli matrix along the $z$ direction. For a two-sublattice antiferromagnetic chain, the exchange field alternates in sign on adjacent sites, $h_i=(-1)^i h$ for $i=2,\ldots,N-1$, while the end sites are taken to be non-magnetic, $h_1=h_N=0$. This choice allows the antiferromagnetic chain to remain magnetically symmetric, while the spin-up and spin-down currents can still differ due to the frequency-induced activation of absorption and emission processes in the nonadiabatic regime. The coefficient $t_{i,i+1}$ specifies the hopping amplitude between sites $i$ and $i+1$ in the chain. It should be noted that two potential barriers of strength $E_b=t_0$ are applied to the two ends of the atomic chain to create an interference region and to observe Fabry-Pérot interferences.

For the electrodes, $d_{j\alpha\sigma}^{\dagger}$ ($d_{j\alpha\sigma}$) are the creation (annihilation) operators for an electron in site j and spin $\sigma$ in lead $\alpha$ (where $\alpha=L$ for the left lead and $\alpha=R$ for the right lead), and $\varepsilon_{j\alpha}$ specifies the corresponding energy. Finally, the term $H_{\mathrm{cpl}}$ describes the tunneling coupling between the end sites of the chain and their corresponding leads. In this relation, $c_{\alpha,\sigma}$ denotes the annihilation operator for an electron with spin $\sigma$ at the end site connected to electrode $\alpha$ (i.e., $c_{L,\sigma}$ corresponds to the first site of the chain and $c_{R,\sigma}$ to the last site). The coefficients $w_\alpha$ describe the tunneling amplitude between the chain and electrode $\alpha$.

The effect of this coupling will be considered in transport calculations through the broadening matrix [31].

To calculate the spin-dependent current through the system, the Keldysh non-equilibrium Green's function (NEGF) formalism was employed [32]. This method, due to its capability to account for non-equilibrium and time-dependent effects, constitutes a powerful tool for studying the spin pumping phenomenon in mesoscopic systems. Within this framework, the retarded and advanced Green's functions between two arbitrary sites $i$ and $j$ of the system are defined as follows, [32]

$$G_{ij,\sigma}^{r}(t,t')=-i\theta(t-t')\langle\{c_{i,\sigma}(t),c_{j,\sigma}^{\dagger}(t')\}\rangle, \tag{5}$$

$$G_{ij,\sigma}^{a}(t,t')=i\theta(t'-t)\langle\{c_{i,\sigma}(t),c_{j,\sigma}^{\dagger}(t')\}\rangle. \tag{6}$$

The Dyson equation for the retarded Green's function in the integral representation is written as,

$$G_{ij}^{r}(t,t') = G_{ij}^{0}(t-t') + \sum\nolimits_{m=1,N} \int dt_1 G_{im}^{r}(t,t_1)\, V_m(t_1)\, G_{mj}^{0}(t_1,t'), \tag{7}$$

where $V_m(t)$ is the time-dependent voltage is applied to positions with coordinates $m = 1, N$, which are connected to the reservoirs. Here, the unperturbed retarded Green's function $G_{m,n}^{0}(t-t')$ describes the central system in the absence of time-dependent potentials. Due to the harmonic nature of external driving, the problem was transformed to the frequency space, and the first harmonic approximation was employed. By applying a partial Fourier transform with respect to the difference time $(t-t')$, the Green's function in the mixed time-frequency space is expressed as,

$$G_{ij}^{r}(t,\omega) = \int_{-\infty}^{t} dt' e^{i\omega(t-t')} G_{ij}^{r}(t,t'). \tag{8}$$

The quantity $G_{ij}^{R}(t,\omega)$ is periodic in time $t$ with period $T_0 = 2\pi/\Omega$ and will be used subsequently to derive the Dyson equation in the Floquet representation. For the retarded Green's function, the Dyson equation can be written as [32],

$$\hat{G}^{r}(t,\omega) = \hat{G}^{0}(\omega) + e^{-i\Omega t}\hat{G}^{r}(t,\omega+\Omega)\,\hat{V}(\Omega)\hat{G}^{0}(\omega) + e^{i\Omega t}\hat{G}^{r}(t,\omega-\Omega)\,\hat{V}(-\Omega)\hat{G}^{0}(\omega). \tag{9}$$

Equation (9) shows how electrons are coupled to a periodic external potential. Here, the full Green's function is obtained by considering the contributions from the unperturbed system as well as the processes involving the energy exchange of quanta $\pm\hbar\Omega$ (photon absorption and emission) during electron transmission through the conductor. Using the Fourier series expansion of the solutions to Eq. (9), we have

$$\hat{G}^{r}(t,\omega) = \sum\nolimits_{m=-k}^{k} \hat{\mathcal{G}}(m,k) e^{-im\Omega t}, \tag{10}$$

which leads to a transformation from the time space to the Floquet-frequency representation, where $\hat{\mathcal{G}}(m,k)$ is the Floquet component of the Green function, defined as

$$\hat{G}(m,\omega) = \frac{1}{T_0}\int_{0}^{T_0} \hat{G}^{R}(t,\omega) e^{im\Omega t} dt. \tag{11}$$

Finally, the current spectral function (i.e., the current per unit energy) can be obtained using the renormalization method in terms of the Floquet components of the retarded Green's function $\hat{G}(m,\omega)$:

$$\jmath_{\alpha}(\omega) = 2Im\big[\hat{\mathcal{G}}(0,\omega)\big]\hat{\Gamma}_{\alpha}(\omega) f_{\alpha}(\omega) + \sum\nolimits_{\beta=L,R}\sum\nolimits_{l=-k}^{k} \hat{\Gamma}_{\alpha}(\omega + l\,\Omega) f_{\beta}(\omega) \hat{\mathcal{G}}(l,\omega)\hat{\Gamma}_{\beta}(\omega)\,\hat{\mathcal{G}}^{\dagger}(l,\omega). \tag{12}$$

In this relation, $\hat{\Gamma}_{\alpha(\beta)}(\omega)$ denotes the hybridization matrix, and $f_{\alpha(\beta)}(\omega)$ is the Fermi distribution function for contact $\alpha(\beta) = L, R$. At the end, for a system driven by a time-periodic potential with period $T_0$, the direct current (DC) component is defined as

$$J_{\text{dc}}^{\alpha} = \frac{1}{2\pi}\int_{-\infty}^{\infty} j_{\alpha}(\omega)\, d\omega, \tag{13}$$

where $j_{\alpha}(\omega) = \text{Tr}[\hat{J}_{\alpha}(\omega)]$.

## 3. Results and discussion

In this section, the results of spin pumping calculations in the antiferromagnetic chain using the Keldysh non-equilibrium Green's function formalism are presented and analysed. In all calculations, the nearest-neighbour hopping parameter $t_0$ measured in units of eV is taken as the energy unit, and all other energy quantities are expressed in terms of it. By reintroducing the reduced Planck constant $\hbar$ in the calculations, the current unit becomes $e[E]/\hbar$, which for $[E]$ in electron volts yields a current on the order of microamperes. In all parts of the calculations, the potential amplitude is $V_p = 0.5$ and the phase difference is $\Delta\phi = \pi/2$.

The effect of the exchange field and driving frequency on the spectral current is examined first. Subsequently, the dependence of the pumped DC current on the chemical potential and driving frequency is analysed, and finally, the possibility of controlling charge and spin currents through external parameters is discussed.

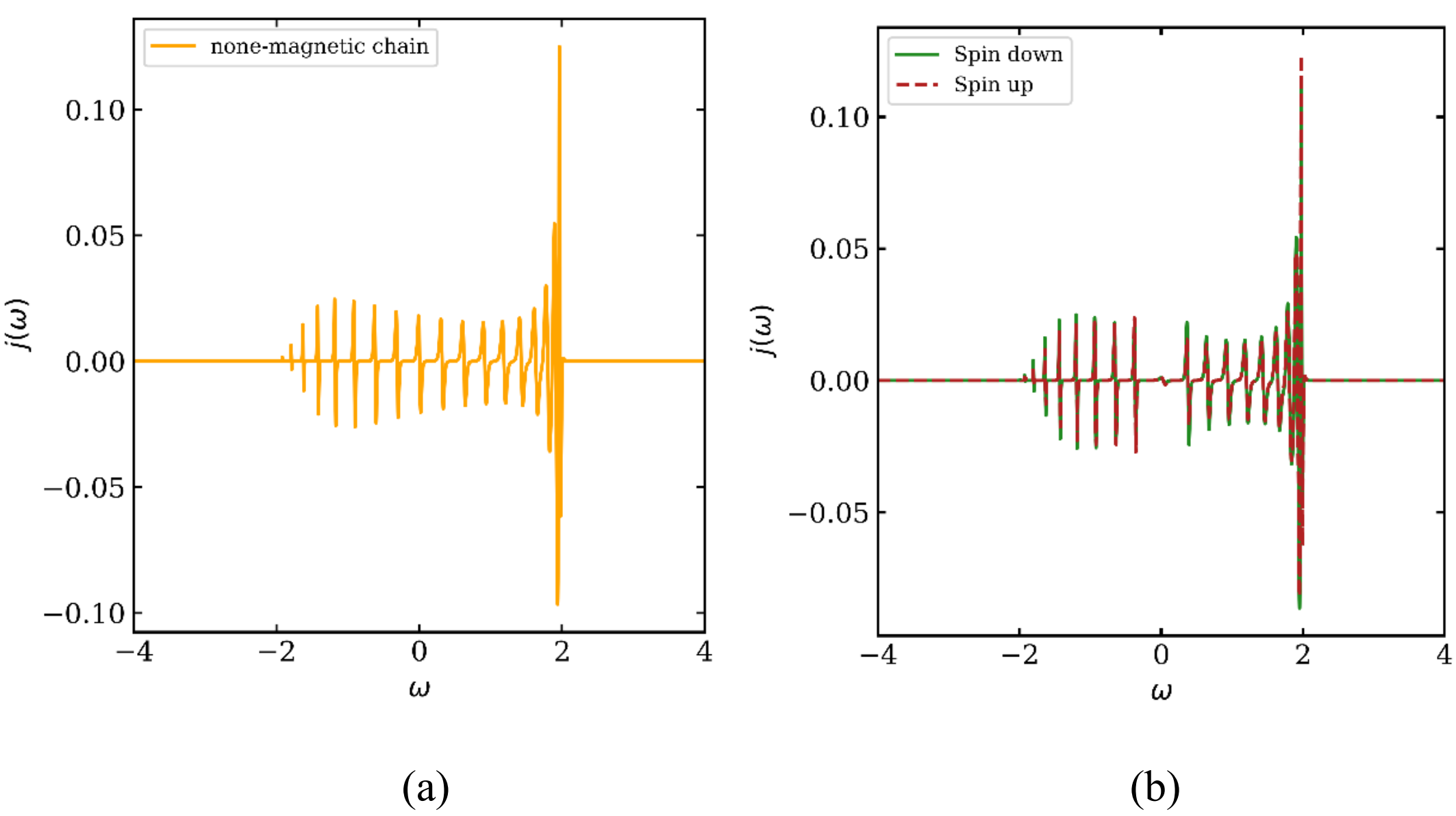

(a) (b)

Figure 2. Spectral current of electrons in (a) non-magnetic and (b) antiferromagnetic configuration with h = 0.2, for driving frequency Ω = 0.01.

Figure 2(a) shows the pumped spectral current at low frequency $\Omega = 0.01$ in the absence of the exchange field $\mathrm{h} = 0$. As observed, the curves corresponding to spin-up and spin-down completely overlap, so that only a single curve is visible in the figure. This complete overlap indicates that in the absence of an exchange field, the system Hamiltonian is symmetric under spin inversion, and the energy structure and transmission probability are identical for both spin channels. Consequently, the spectral current for both spins is the same across the entire energy range, and no spin polarization or separation is generated in the system. On the other hand, the spectral current is non-zero only within a limited energy window and approaches zero outside this range, indicating that a limited number of energy states participate in the quantum pumping process in the adiabatic regime.

In Figure 2(b), the same calculations are performed in the presence of an exchange field of strength $\mathrm{h} = 0.2$. The most significant change compared to Figure 2(a) is the opening of an energy gap around the Fermi energy, which eliminates the spectral current contribution in this region. This behavior indicates that the exchange field modifies the electronic structure of the antiferromagnetic chain and, by creating a gap, restricts electron access to certain transmission states. Despite this change, the curves corresponding to spin-up and spin-down remain nearly overlapping, and no significant difference between the responses of the two spin channels is observed. Thus, in the adiabatic regime, the dominant effect of the exchange field is to create a gap in the energy structure, while the transmission response of the system remains approximately identical for both spin directions.

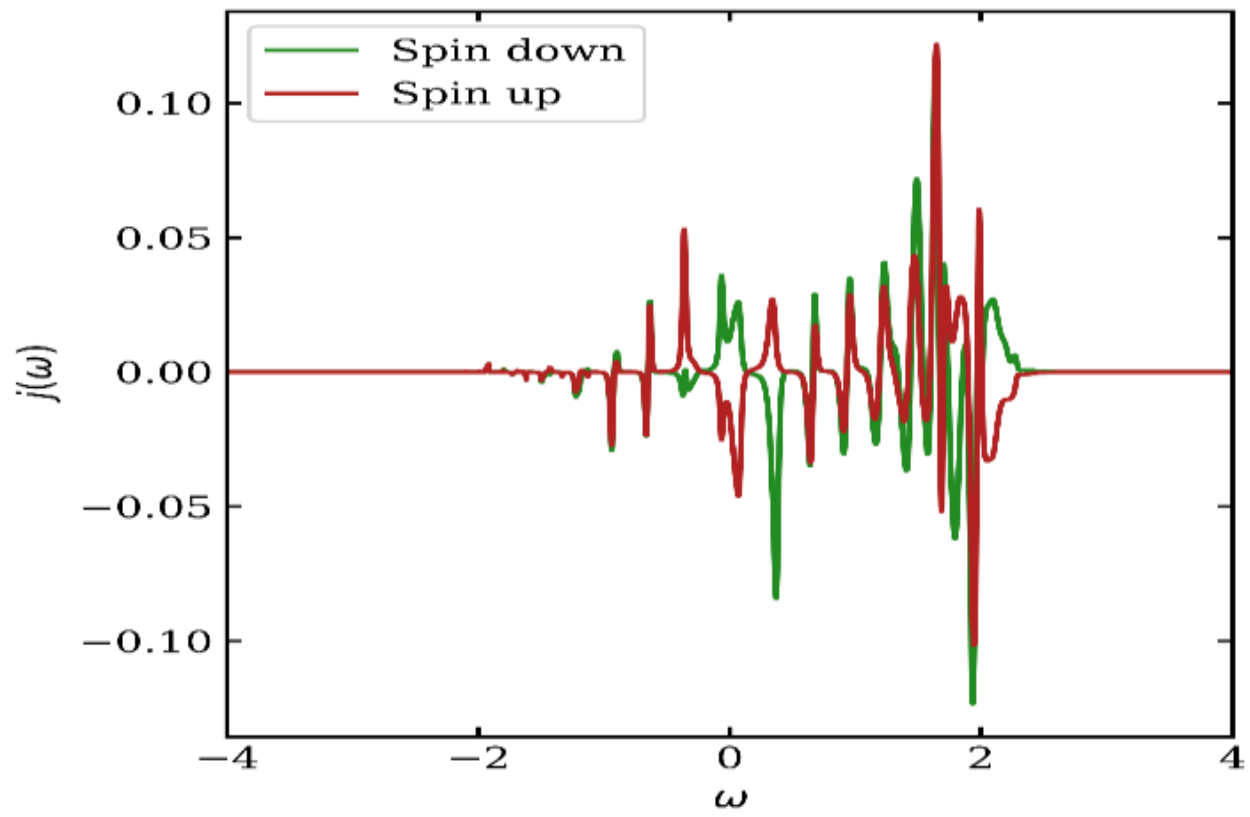

Figure 3. Spectral current for spin-up (red) and spin-down (green) in the presence of exchange field $\mathrm{h} = 0.2$ and driving frequency $\Omega = 0.3$.

Figure 3 displays the spectral current for spin-up and spin-down in the presence of exchange field $\mathrm{h} = 0.2$ and frequency $\Omega = 0.3$. Other parameters are the same as chosen in Figure 2.

Compared to the adiabatic regime, the spectral structure of both spin components becomes significantly more complex. The number of oscillations increases, and the spectral current is more broadened compared to the low driving frequency case (see Fig. 2(b)). Furthermore, multiple positive and negative regions appear in the spectrum, indicating competition among different transmission channels in the presence of a high-frequency periodic drive. This behaviour shows that as the system enters the nonadiabatic regime, absorption and emission processes of energy quanta $\hbar\Omega$ become activated, and the contribution of higher-order transmission channels to the pumping process increases. Moreover, comparison of the two curves reveals that the transmission response for spin-up and spin-down is no longer identical. Although both components exhibit similar oscillatory structures, noticeable differences are observed in the positions of some peaks, their amplitudes, and the distribution of positive and negative regions. This difference indicates that in the presence of the exchange field, increasing the driving frequency causes the nonadiabatic processes to act differently for the two spin channels. Consequently, the transmission probability, resonance conditions, and transmission phase for spin-up and spin-down electrons, giving rise to a spin-dependent transmission response. This spectral separation provides the basis for generating a pure spin current in the absence of external bias, which will be examined further.

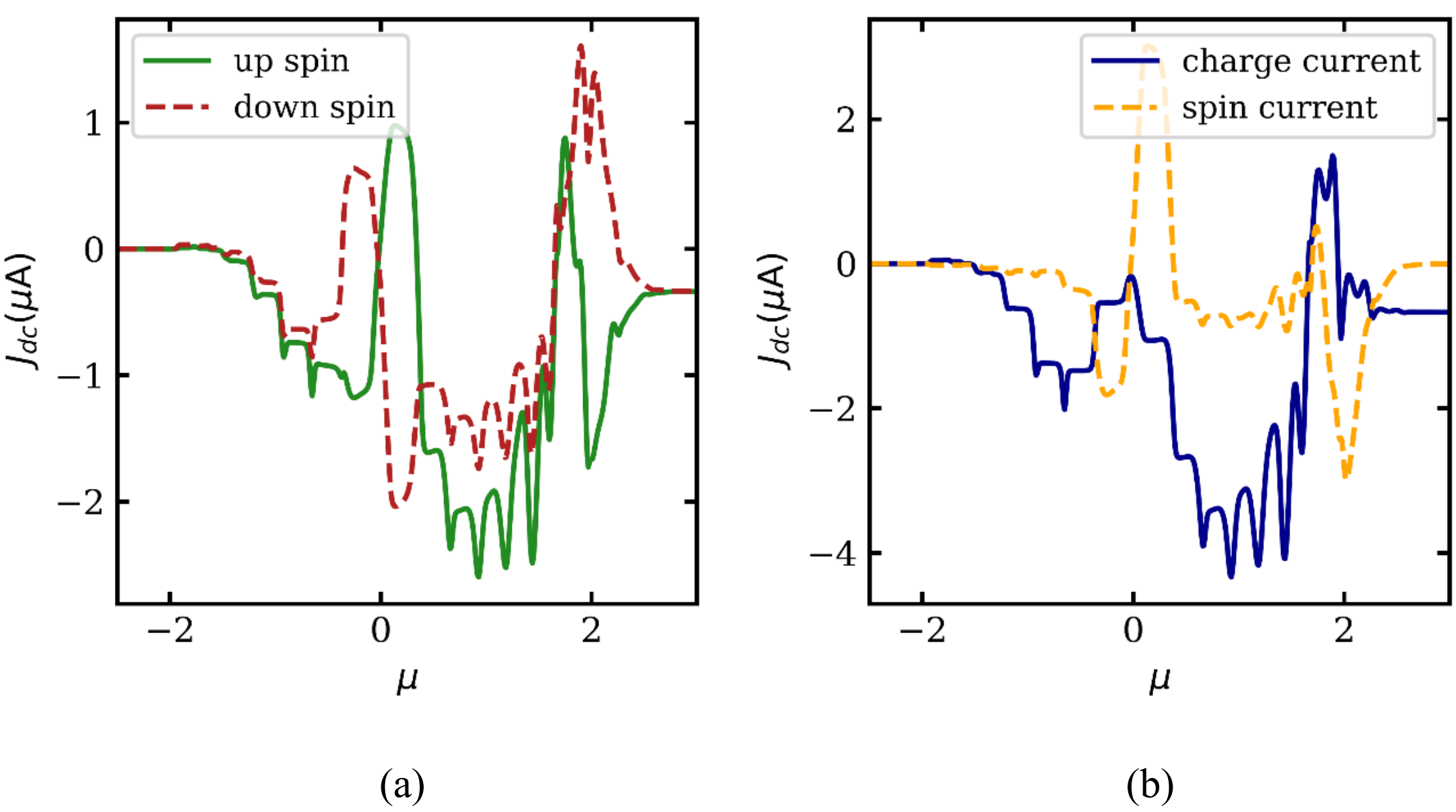

Figure 4. (a) Pumped DC current for spin-up (solid green line) and spin-down (dashed red line) as a function of chemical potential $\mu$, in the presence of exchange field $\mathrm{h} = 0.2$ and driving frequency $\Omega = 0.3$. (b) Charge current ($I_c = I_\uparrow + I_\downarrow$) (solid green line) and spin current ($I_s = I_\uparrow - I_\downarrow$) (dashed red line) as a function of chemical potential.

Figure 4(a) shows the pumped DC current for spin-up and spin-down as a function of chemical potential in the presence of exchange field with $\mathrm{h} = 0.2$ and frequency $\Omega = 0.3$. Other parameters are chosen as in Figure 2. It is observed that the DC current exhibits a strong dependence on the chemical potential position, and with variation of $\mu$, both spin components show significant changes in magnitude and direction of the current. This behaviour indicates that the quantum pumping process is strongly influenced by the available electronic states, and as the chemical potential shifts, the contribution of different transmission channels to the pumping process changes.

In regions far from the center of the spectrum, the current for both spins is very small and no significant difference between them is observed. As the chemical potential approaches the allowed energy region, the current increases rapidly and a series of consecutive peaks and valleys appears. This oscillatory structure indicates the occurrence of resonance conditions for different transmission channels and the variation of their contribution to the total current. A notable point in this figure is the clear difference in the response of the two spin components in different ranges of chemical potential. Although in some regions the current magnitudes for the two spins are approximately similar, in many energy ranges a significant difference in amplitude and even sign of the current is observed. For example, around $\mu \cong 0$, the spin-up current has a positive value, while the spin-down current reaches significant negative values. Also, in the range $\mu \cong 1.5 - 2$, both currents undergo strong variations, but the positions of the peaks and their maximum values are not the same for the two spins. This difference indicates that the exchange field combined with nonadiabatic driving changes the resonance conditions for the two spin channels differently, resulting in a spin-dependent transmission response.

From a physical perspective, changing the chemical potential shifts the Fermi energy relative to the band structure of the system. Consequently, the transmission channels participating in the pumping process continuously change. Since in the nonadiabatic regime these channels are also affected by absorption and emission of energy quanta, the competition among them gives rise to peaks, current sign reversal, and significant differences between the two spin components.

Overall, these results indicate that the chemical potential, in addition to the driving frequency, is an effective control parameter for tuning the magnitude and direction of the pumped current as well as the degree of spin separation. Therefore, by appropriately adjusting the Fermi energy, regions can be selected where the difference between spin-up and spin-down currents is maximized, and the system's performance as a spin pump is improved.

To evaluate the system's performance as both a charge pump and a spin pump, Figure 4(b) presents the charge current ($I_c = I_\uparrow + I_\downarrow$) and the spin current ($I_s = I_\uparrow - I_\downarrow$) as function of chemical potential. It is observed that both quantities exhibit a strong dependence on the chemical potential, with significant changes in magnitude and even direction as $\mu$ varies. This behaviour indicates that shifting the Fermi energy alters the contribution of different transmission states to the pumping process, thereby continuously redistributing the charge and spin current contributions.

As it is demonstrated in the figure the maxima of the charge and spin currents do not occur at the same chemical potentials, and in some chemical potential ranges, one of them has a significant value while the other is relatively small. This difference indicates that optimal conditions for charge transfer are not necessarily the same as those for spin transfer. In other words, by appropriate selection of the chemical potential, the performance of the system can be directed toward predominance of either charge or spin transfer.

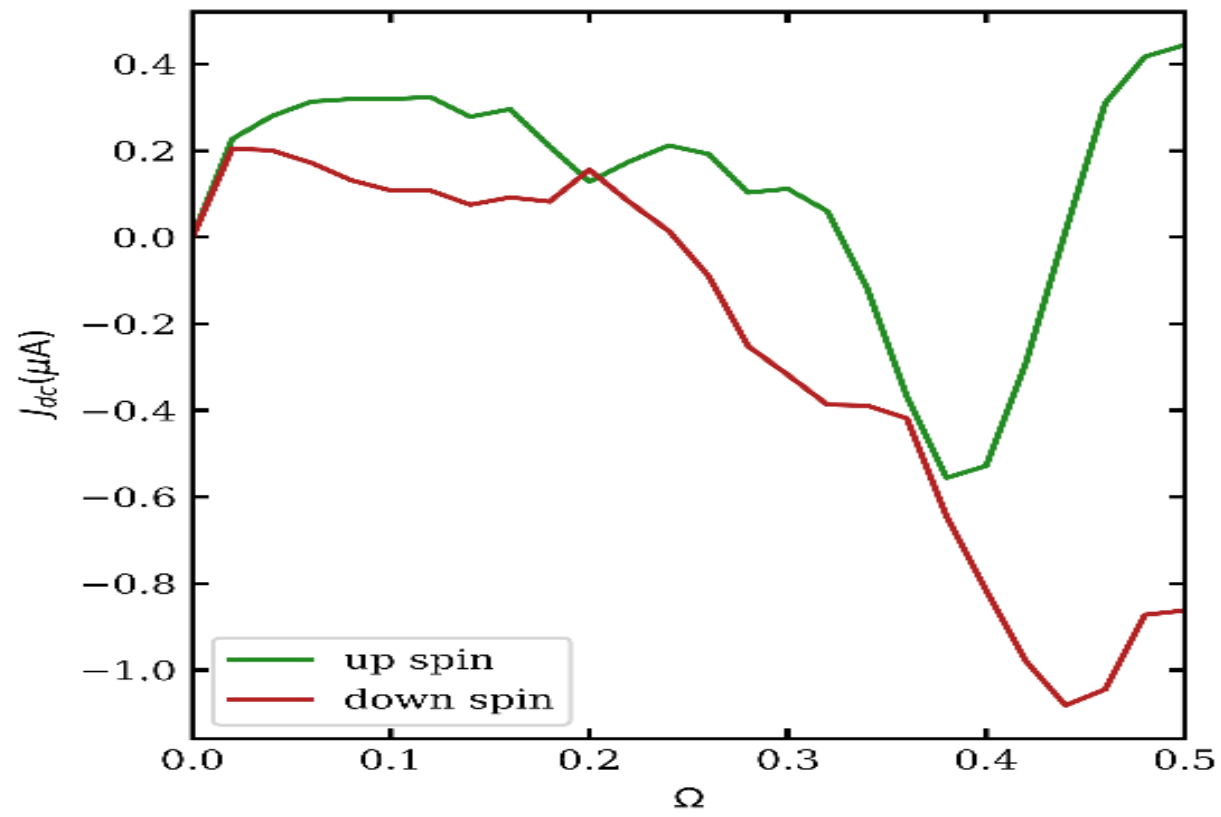

Figure 5. Pumped DC current for spin-up (green) and spin-down (red) as a function of driving frequency Ω, at chemical potential $\mu = 0$, and exchange field h $= 0.2$.

To investigate the transition from the adiabatic to the nonadiabatic regime, Figure 5 shows the pumped DC current for spin-up and spin-down as a function of driving frequency, with chemical potential $\mu = 0$. Other parameters are chosen as in Figure 2. At low frequencies, both currents have positive values and the difference between them is negligible, consistent with the nearly identical behaviour of spectral currents in Figure 2(b). In this regime, although the exchange field creates an energy gap, it does not yet significantly separate the transmission responses of the two spin channels.

With increasing the frequency, the difference between the two spin currents gradually enhances. The spin-down current initially decreases and then changes sign, while the spin-up current remains positive up to higher frequencies and experiences a significant reduction only over a limited range.

This behaviour results from the rearrangement of spectral currents observed in Figure 3; the activation of nonadiabatic transmission channels causes the positive and negative contributions of the spectral current to change differently for the two spins. Since the DC current is obtained by integrating the spectral current over energy, this difference in spectral distribution directly leads to significant differences in the magnitude and even direction of the DC current for the two spin components.

Overall, the results of this figure show that the exchange field alone mainly modifies the energy structure, but its combination with nonadiabatic driving creates effective separation of spin currents. This finding indicates that the driving frequency can serve as an effective control parameter for tuning the spin response of the system and designing spin-pumping devices based on antiferromagnetic chains.

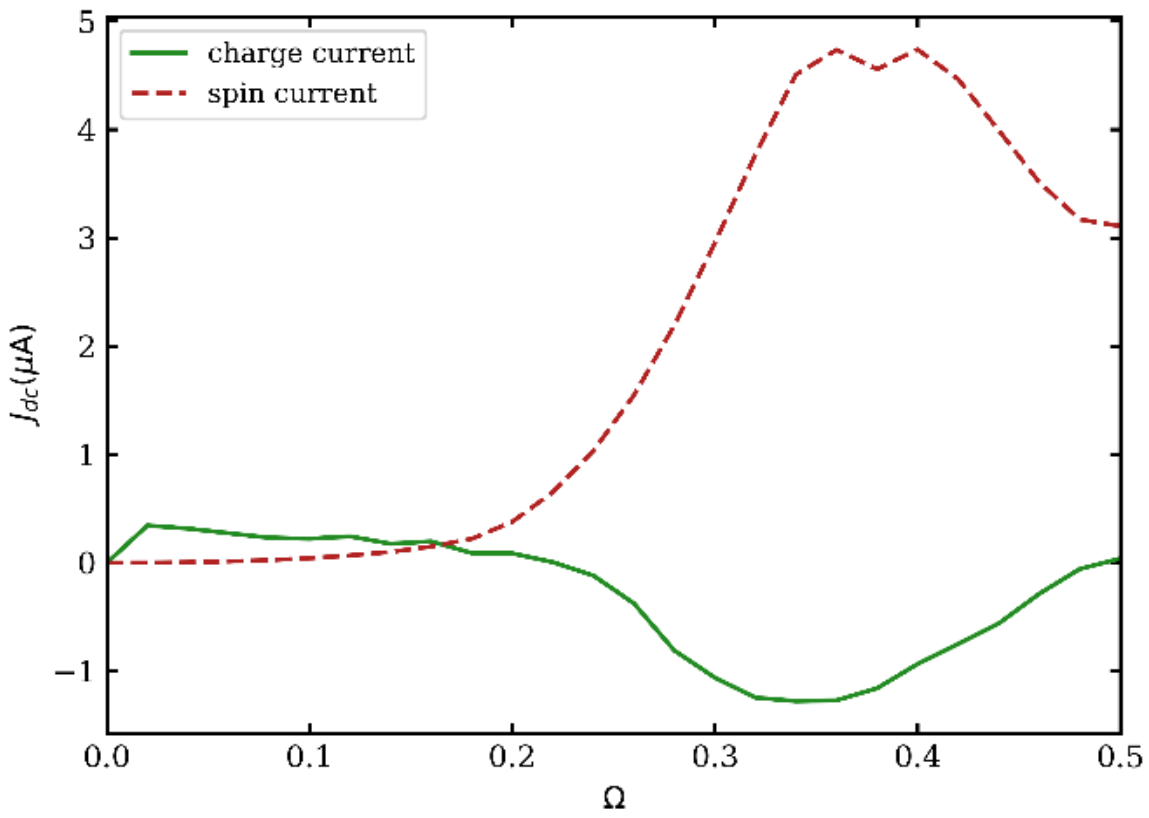

Figure 6. Dependence of charge current (solid green line) and spin current (dashed red line) on driving frequency Ω, in the presence of exchange field $h = 0.2$, and chemical potential $\mu = 0.2$.

Figure 6 shows the dependence of the charge and spin currents on the driving frequency in the presence of the exchange field for $\mu = 0.2$ and the exchange field $h = 0.2$. At low frequencies, both currents have small values, characteristic of the adiabatic regime. As the frequency increases, the spin current significantly enhances, while the charge current, after changing sign, acquires a smaller magnitude compared to the spin current over a wide frequency range. A notable point is that around $\Omega \cong 0.5$, the charge current nearly vanishes, while the spin current still maintains a considerable value. This behaviour indicates that in this region, the contributions of spin-up and spin-down currents almost cancel each other in terms of charge, but their difference remains non-zero. Consequently, the system can generate a significant spin current without net charge transfer. This feature demonstrates that by proper selection of control parameters such as chemical potential and driving frequency, nearly pure spin pumping can be achieved in the antiferromagnetic chain.

**4. Conclusion**

In this work, we have theoretically investigated quantum spin pumping in an antiferromagnetic chain driven by time-dependent periodic potentials, using the Keldysh non-equilibrium Green's function formalism. Our study demonstrates that pure spin currents can be generated and controlled in the absence of any external bias by appropriate tuning of system parameters, including the exchange field, driving frequency, and chemical potential.

The results reveal that in the adiabatic regime (low driving frequencies), the spectral currents for spin-up and spin-down electrons remain nearly identical, with the exchange field primarily modifying the electronic structure by opening an energy gap around the Fermi level without inducing significant spin separation. However, as the driving frequency increases and the system enters the nonadiabatic regime, absorption and emission processes of energy quanta become activated, leading to distinct transmission responses for the two spin channels. This frequency-induced symmetry breaking results in substantial differences in both the magnitude and direction of spin-up and spin-down pumped currents.

Furthermore, we have shown that the chemical potential serves as an effective control parameter for tuning the charge and spin currents. The pumped currents exhibit strong dependence on the Fermi energy, with pronounced peaks, valleys, and sign reversals arising from the competition among different transmission channels. Notably, we have found that the maxima of charge and spin currents occur at different chemical potentials, enabling selective optimization for either charge or spin transport.

One of the most significant findings of this work is the demonstration that nearly pure spin pumping can be achieved in the antiferromagnetic chain. By selecting appropriate values of the chemical potential and driving frequency, the charge current can be substantially suppressed while maintaining a considerable spin current. This capability is particularly relevant for low-power spintronic applications, as it enables the generation of spin currents with minimal energy dissipation and Joule heating.

Our results provide a comprehensive understanding of quantum spin pumping in antiferromagnetic systems and highlight the key role of nonadiabatic dynamics in achieving efficient spin current generation. The ability to electrically control both the magnitude and direction of spin currents through chemical potential tuning, without the need for external magnetic fields or ferromagnetic electrodes, offers promising prospects for designing next-generation spintronic devices based on antiferromagnetic materials. These findings may pave the way for the development of energy-efficient, high-speed spin-based information processing technologies.